## ARTICLE INFORMATION

**Article title**

4D-CTA Image and geometry dataset for kinematic analysis of abdominal aortic aneurysms

**Authors**

Mostafa Jamshidian[1]*, Adam Wittek[1], Saeideh Sekhavat[1], Farah Alkhatib[2], Jens Carsten Ritter[3,4], Paul M. Parizel[5,6], Donatien Le Liepvre[7], Florian Bernard[7], Ludovic Minvielle[7], Antoine Fondanèche[7], Jane Polce[8], Christopher Wood[8], Karol Miller[1]

**Affiliations**

[1]Intelligent Systems for Medicine Laboratory, The University of Western Australia, Perth, Western Australia, Australia

[2]Department of Mechanical Engineering, The University of Western Australia, Perth, Western Australia, Australia

[3]Department of Vascular Surgery, Fiona Stanley Hospital, Perth, Australia

[4]Curtin University, School of Medicine, Perth, Australia

[5]Department of Diagnostic and Interventional Radiology, Royal Perth Hospital, Perth, Western Australia, Australia

[6]Medical School, University of Western Australia (UWA), Perth, Western Australia, Australia

[7]Nurea, Bordeaux, France

[8]Department of Radiology, Fiona Stanley Hospital, Perth, Australia

**Corresponding author's email address and Twitter handle**

mostafa.jamshidian@uwa.edu.au
@MostafaJAU

**Abstract**

This article presents a dataset used in the article "Kinematics of Abdominal Aortic Aneurysms" [1], published in the Journal of Biomechanics. The dataset is publicly available for download from the Zenodo data repository (https://doi.org/10.5281/zenodo.15477710). The dataset includes time-resolved 3D computed tomography angiography (4D-CTA) images of abdominal aortic aneurysm (AAA) captured throughout the cardiac cycle from ten patients diagnosed with AAA, along with ten patient-specific AAA geometries extracted from these images. Typically, the 4D-CTA dataset for each patient contains ten electrocardiogram (ECG)-gated 3D-CTA image frames acquired over a cardiac cycle, capturing both the systolic and diastolic phases of the AAA configuration. For method verification, the dataset also includes synthetic ground truth data generated from Patient 1's 3D-CTA AAA image in the diastolic phase. The ground truth data includes the patient-specific finite element (FE) biomechanical model and a synthetic systolic 3D-CTA image. The synthetic systolic image was generated by warping Patient 1's diastolic 3D-CTA image using the realistic displacement field obtained from the AAA biomechanical FE model. The images were acquired at Fiona Stanley Hospital in Western Australia and provided to the researchers at the Intelligent Systems for Medicine Laboratory at The University of Western Australia (ISML-UWA), where image-based AAA kinematic analysis was performed using a newly created algorithm, as described in [1]. The AAA geometries were extracted using an automated image processing pipeline comprising AI-based segmentation with PRAEVAorta software by NUREA (https://www.nurea-soft.com/), automated post-processing with the ISML-UWA in-house code (https://arxiv.org/abs/2403.07238), and surface model extraction using the freely available BioPARR (Biomechanics-based Prediction of Aneurysm Rupture Risk) (https://bioparr.mech.uwa.edu.au/) and 3D Slicer (https://www.slicer.org/) software packages [2,3]. Our dataset enabled the analysis of AAA wall displacement and strain throughout the cardiac cycle using a non-invasive, in vivo, image registration-based approach [1]. The use of widely adopted, open-source file formats—NRRD for images and STL for geometries—facilitates broad applicability and reusability in AAA biomechanics studies that require patient-specific geometry and information about AAA kinematics during cardiac cycle.

# SPECIFICATIONS TABLE

| Subject | Engineering & Materials science |
|---|---|
| Specific subject area | Computational Biomechanics and Kinematic Analysis of Abdominal Aortic Aneurysms |
| Type of data | *Data for ten patients:*<br>4D-CTA images: Nearly Raw Raster Data file (*.nrrd)<br>AAA external wall geometries: STL unstructured triangulated surface file (*.stl)<br><br>*Ground truth data for method verification (generated from Patient 1 data):*<br>AAA surface model: STL unstructured triangulated surface file (*.stl)<br>Finite element model: Abaqus FE model file (*.inp)<br>Finite element mesh: Abaqus FE mesh file (*.inp)<br>FE nodes displacement field: Visualization Toolkit (VTK) files (*.vtu)<br>FE nodes coordinates: Microsoft Excel Comma Separated Values File (*.csv)<br>Image warping transform: 3D Slicer HDF5 transform file (*.h5)<br>3D-CTA synthetic image: Nearly Raw Raster Data file (*.nrrd) |
| Data collection | Time-resolved 3D computed tomography angiography (4D-CTA) images were acquired using a Siemens SOMATOM Definition Flash CT Scanner. For each patient, the 4D-CTA dataset typically includes ten electrocardiogram ECG-gated 3D-CTA image frames captured uniformly at 10% intervals of the R–R interval across the cardiac cycle. The R–R interval represents the time between two successive R- |
| | waves (peak points) of the ECG signal. Geometries were extracted from the medical images using an automated image processing pipeline comprising AI-based segmentation with PRAEVAorta by NUREA (https://www.nurea-soft.com/), automated post-processing with in-house MATLAB code (https://arxiv.org/abs/2403.07238), and surface model extraction using the BioPARR (Biomechanics-based Prediction of Aneurysm Rupture Risk) (https://bioparr.mech.uwa.edu.au/) and 3DSlicer (https://www.slicer.org/) software packages [2,3].<br><br>For generating the ground truth data, we used the AAA surface model of Patient 1 created using BioPARR. HyperMesh software (https://altair.com/hypermesh) was used to mesh the AAA model with hexahedral elements, and the mesh was then imported into Abaqus software (https://www.3ds.com/products/simulia/abaqus). The FE model solution, obtained using the Abaqus/Standard solver, outputs the displacements of the FE nodes. We used the nodal coordinates of the AAA FE mesh in undeformed and deformed configurations to create the image warping transform using the Scattered Transform Extension [4] in 3D Slicer [3]. Then, using the 3D Slicer Transforms module, we applied the transform to Patient 1's 3D-CTA image in the diastolic phase and generated the synthetic systolic 3D-CTA image, with the known AAA wall displacement, obtained from the biomechanical FE model, as ground truth. |
| **Data source location** | Images were acquired at Fiona Stanley Hospital, Perth, Western Australia, Australia. The images are stored at Fiona Stanley Hospital. The geometries and ground truth data are stored at The University of Western Australia, Perth, Western Australia, Australia. |
| **Data accessibility** | Repository name: Zenodo<br><br>Data identification number: 10.5281/zenodo.15477710<br><br>Direct URL to data: https://doi.org/10.5281/zenodo.15477710 |
| **Related research article** | M. Jamshidian, A. Wittek, S. Sekhavat, K. Miller, Kinematics of abdominal aortic Aneurysms, J Biomech 179 (2025) 112484. https://doi.org/10.1016/j.jbiomech.2024.112484 |

# VALUE OF THE DATA

- The dataset described in this paper was used in our recently published study on the kinematics of abdominal aortic aneurysm (AAA) [1]. The presented dataset can be used to reproduce the results of that study and serve as a benchmark for evaluating alternative methods in future research. The ground truth data included in the dataset may be used to verify new methods.
- The medical images and geometries described in this paper provide a valuable resource for researchers developing new computational methods for medical image analysis and computational biomechanics. The dataset supports research focused on non-invasive, in vivo, image-based measurement of AAA wall strain, as well as broader application of computational biomechanics techniques for AAA analysis.
- By providing the images and geometries in standard, widely used file formats compatible with open-source software, we aim to enable broader reuse of the dataset not only by researchers focusing on AAA biomechanics but also by those conducting related studies in medical image analysis and computational biomechanics using various methods and platforms.

# BACKGROUND

Abdominal aortic aneurysm (AAA), characterized by permanent and irreversible dilation of the aorta, is often asymptomatic but fatal in most cases if ruptured [5]. Current disease management relies on AAA maximum diameter and growth rate criteria [5], which may not accurately reflect rupture risk in individual patients. Biomechanical analysis offers a more personalized approach to AAA risk assessment beyond the limitations of the maximum diameter criteria [6]. In this context, while AAA wall stress has been extensively studied, AAA kinematics has received relatively limited attention [1]. This underexplored area has the potential to provide complementary insights into disease progression and rupture risk. Motivated by this gap, we compiled a dataset to support the development and validation of a novel image-based, non-invasive method for AAA kinematic analysis using time-resolved 3D computed tomography angiography (4D-CTA) images [1]. To the best of our knowledge, this dataset is the first publicly available 4D-CTA resource for AAA.

The present data article complements a previously published research article [1] by providing open access to the medical images, geometries, and synthetic ground truth used in method development and verification. This allows researchers to reproduce the results of that study or use the data to develop alternative methods and evaluate the accuracy of new approaches against the ground truth data.

# DATA DESCRIPTION

We provide medical images and surface geometries for ten AAA patients analysed in Jamshidian et al. [1], along with the ground truth data used for method verification therein. A subset of medical images from this dataset has been also used in the study on methods for generation of patient-specific computational finite element (FE) meshes of AAA by Alkhatib et al. [7]. As illustrated in Fig. 1, the dataset is organized into 11 folders: ten patient folders labelled P1 through P10, containing patient images and geometry, and one folder labelled "Ground Truth," which contains data for method verification.

Each of the ten patient folders (P1 to P10) contains cropped 3D-CTA images of the patient's AAA at various phases of the cardiac cycle. The image files names — 10%.nrrd, 20%.nrrd, 30%.nrrd, up to 100%.nrrd — indicate the corresponding phase of the cardiac cycle. For example, 30%.nrrd represents the 30% phase.

The Number of images acquired across the cardiac cycle varied between patients. Therefore, folders for patients 3, 5, 6, 7, 9, and 10 contain images from 10 phases (10% to 100%). Folders for patients 1 and 8 contain images from 7 phases (30% to 90%). Folders for patients 2 and 4 contain images from only 2 phases (40% and 80%). Overall, images from the 10%, 20%, and 100% phases were generally of lower quality compared to those from the 30% to 90% phases.

In the image registration-based AAA kinematics analysis conducted by Jamshidian et al. [1], the 30% phase image was used as the AAA systolic image and the 90% phase image as the AAA diastolic image for patients 1, 2, 5, 6, 7, 8, 9, and 10. Since the 30% and 90% phase images were not available for patients 2 and 4, the 40% phase image was used as the AAA systolic image and the 80% phase image as the AAA diastolic image.

In addition to the 3D-CTA images, each patient folder includes a triangulated surface model of the patient's AAA external wall in the systolic configuration, provided as Wall.stl. The surface model is defined in the patient coordinate system RAS, aligned with the anatomical axes: left–right (R), posterior–anterior (A), and inferior–superior (S).

As an example, Fig. 2 shows the cropped 3D-CTA image of patient 1's AAA in the systolic phase (30% of the cardiac cycle), and the surface model of the AAA external wall.

In addition to the folders containing images and geometries for 10 patients, the dataset also includes a "Ground Truth" folder (shown in Fig. 1), which contains data used to verify the AAA kinematic analysis method developed and employed by Jamshidian et al. [1]. The contents of this folder are:

- CTA_undeformed.nrrd – Undeformed (diastolic) 3D-CTA image of the AAA
- CTA_deformed.nrrd – Deformed (synthetic systolic) 3D-CTA image of the AAA
- FE_model.inp – input file for patient-specific FE biomechanical model in Abaqus finite element code format
- FE_mesh_undeformed.inp – FE mesh for undeformed (diastolic) geometry in Abaqus finite element code format
- FE_mesh_deformed.inp – FE mesh for deformed (synthetic systolic) geometry in Abaqus finite element code format
- FE_nodes_undeformed_coordinates.csv – Nodal coordinates of the FE mesh for undeformed (diastolic) configuration
- FE_nodes_deformed_coordinates.csv – Nodal coordinates of the FE mesh for deformed (synthetic systolic) configuration
- FE_nodes_displacement.vtu – Nodal displacements obtained from the patient-specific FE biomechanical model used to generate the synthetic systolic AAA geometry
- Transform.h5 – Image warping transform used to deform the undeformed (diastolic) 3D-CTA image of Patient 1
- Wall.stl – Triangulated surface model of the AAA external wall
- AAA.stl – Triangulated surface model of the AAA

Files in the "Ground Truth" folder enable researchers to either regenerate the ground truth data or directly use it to evaluate new methods based on known AAA wall displacements. Fig. 3 shows the triangulated surface model of the AAA, the FE mesh, and a contour plot of the AAA wall displacements obtained from the FE biomechanical model used to generate the synthetic systolic geometry and image.

| Dataset folder | Ground Truth folder |
|---|---|
| 📁 Ground Truth | 📄 Transform.h5 |
| 📁 P1 | 📄 FE_mesh_deformed.inp |
| 📁 P2 | 📄 FE_mesh_undeformed.inp |
| 📁 P3 | 📄 FE_model.inp |
| 📁 P4 | 📄 FE_nodes_deformed_coordinates.csv |
| 📁 P5 | 📄 FE_nodes_undeformed_coordinates.csv |
| 📁 P6 | 📄 CTA_deformed.nrrd |
| 📁 P7 | 📄 CTA_undeformed.nrrd |
| 📁 P8 | 📄 AAA.stl |
| 📁 P9 | 📄 Wall.stl |
| 📁 P10 | 📄 FE_nodes_displacement.vtu |

| Folder P1 | Folder P2 | Folder P3 | Folder P4 | Folder P5 |
|---|---|---|---|---|
| 30%.nrrd | 40%.nrrd | 10%.nrrd | 40%.nrrd | 10%.nrrd |
| 40%.nrrd | 80%.nrrd | 20%.nrrd | 80%.nrrd | 20%.nrrd |
| 50%.nrrd | Wall.stl | 30%.nrrd | Wall.stl | 30%.nrrd |
| 60%.nrrd | | 40%.nrrd | | 40%.nrrd |
| 70%.nrrd | | 50%.nrrd | | 50%.nrrd |
| 80%.nrrd | | 60%.nrrd | | 60%.nrrd |
| 90%.nrrd | | 70%.nrrd | | 70%.nrrd |
| Wall.stl | | 80%.nrrd | | 80%.nrrd |
| | | 90%.nrrd | | 90%.nrrd |
| | | 100%.nrrd | | 100%.nrrd |
| | | Wall.stl | | Wall.stl |

| Folder P6 | Folder P7 | Folder P8 | Folder P9 | Folder P10 |
|---|---|---|---|---|
| 10%.nrrd | 10%.nrrd | 30%.nrrd | 10%.nrrd | 10%.nrrd |
| 20%.nrrd | 20%.nrrd | 40%.nrrd | 20%.nrrd | 20%.nrrd |
| 30%.nrrd | 30%.nrrd | 50%.nrrd | 30%.nrrd | 30%.nrrd |
| 40%.nrrd | 40%.nrrd | 60%.nrrd | 40%.nrrd | 40%.nrrd |
| 50%.nrrd | 50%.nrrd | 70%.nrrd | 50%.nrrd | 50%.nrrd |
| 60%.nrrd | 60%.nrrd | 80%.nrrd | 60%.nrrd | 60%.nrrd |
| 70%.nrrd | 70%.nrrd | 90%.nrrd | 70%.nrrd | 70%.nrrd |
| 80%.nrrd | 80%.nrrd | Wall.stl | 80%.nrrd | 80%.nrrd |
| 90%.nrrd | 90%.nrrd | | 90%.nrrd | 90%.nrrd |
| 100%.nrrd | 100%.nrrd | | 100%.nrrd | 100%.nrrd |
| Wall.stl | Wall.stl | | Wall.stl | Wall.stl |

**Fig. 1.** Overview of the dataset directory structure and contents, consisting of ten patient folders (P1 to P10) and one "Ground Truth" folder. Each patient folder contains cropped 3D-CTA images of the patient's abdominal aortic aneurysm (AAA) at multiple cardiac phases (e.g., 30%.nrrd, representing 30% of the cardiac cycle) and a triangulated surface model (Wall.stl) of the AAA external wall in the systolic phase. The "Ground Truth" folder includes data used to verify the AAA kinematic analysis method developed by Jamshidian et al. [1]. These data allow researchers to reproduce the ground truth or use it directly to assess the accuracy of new AAA kinematic analysis methods.

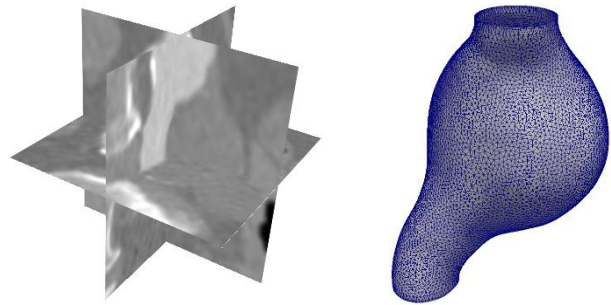

**Fig. 2.** Example visualization of Patient 1's AAA dataset, showing the cropped 3D-CTA image at the 30% phase of the cardiac cycle (systolic phase), alongside the corresponding triangulated surface model of the AAA external wall. The geometry is provided as an STL file in the patient's RAS coordinate system, aligned with the anatomical axes: left–right (R), posterior–anterior (A), and inferior–superior (S).

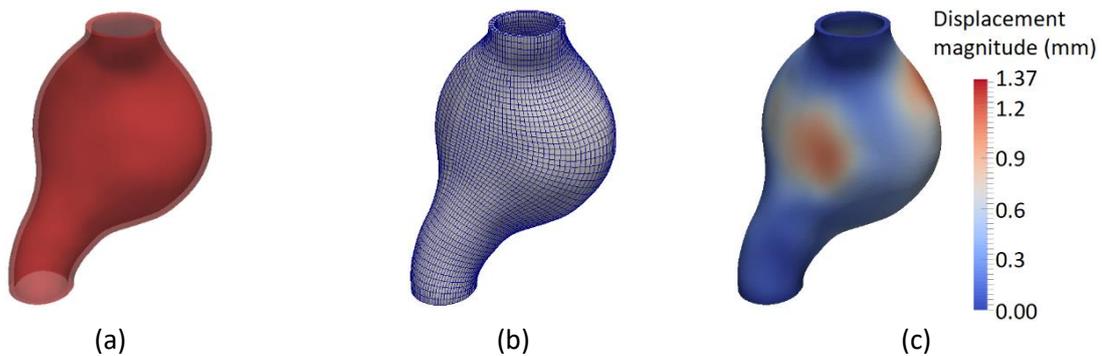

(a)　　　　　　　　　(b)　　　　　　　　　(c)

**Fig. 3.** Visualization of the synthetic ground truth data used for verifying AAA kinematic analysis methods: (a) The triangulated surface model of Patient 1's AAA with an assumed wall thickness. (b) The hexahedral finite element (FE) mesh created in HyperMesh (https://altair.com/hypermesh) and imported into the Abaqus FE software (https://www.3ds.com/products/simulia/abaqus) for patient-specific biomechanical modelling. (c) The resulting displacement of the AAA wall from the FE biomechanical simulation, used to warp the diastolic-phase 3D-CTA image of Patient 1's AAA and generate a synthetic systolic image. This synthetic ground truth supports method verification in line with FDA and ASME guidelines by providing physiologically plausible AAA wall displacement.

## EXPERIMENTAL DESIGN, MATERIALS AND METHODS

### Image

Although 4D-CT is not routinely used for AAA diagnosis or monitoring, it is increasingly employed in the diagnosis of cardiac diseases and can be performed using scanners commonly available in hospital radiology departments [8]. Therefore, image acquisition in this study did not require any specialized equipment beyond standard clinical CT scanners. Contrast-enhanced 4D-CTA images of the abdominal aorta, synchronized with the cardiac cycle, were acquired using electrocardiogram (ECG) gating on a Siemens SOMATOM Definition Flash CT scanner (Siemens Healthineers AG, Forchheim, Germany). Typically, ten 3D image volumes were captured at uniformly sampled time points throughout the cardiac cycle.

**Segmentation**

To create a patient-specific AAA surface geometry model from the patient's 3D-CTA image in the systolic phase, we used an automated pipeline consisting of (1) AI-based segmentation using the PRAEVAorta software by NUREA (https://www.nurea-soft.com/), (2) automated post-processing with in-house MATLAB code (https://arxiv.org/abs/2403.07238), and (3) extraction of the surface model using the open-source BioPARR (Biomechanics-based Prediction of Aneurysm Rupture Risk) (https://bioparr.mech.uwa.edu.au/) software package [2].

PRAEVAorta is an AI-based decision support software developed by NUREA that performs automated reconstruction of the arterial tree from medical images using advanced image processing algorithms and numerical simulation tools. In this study, we used the PRAEVAorta Web platform to segment the 3D-CTA images. The output included three labelled (segmented) regions: (1) the aneurysm wall and intraluminal thrombus (ILT), (2) calcification, and (3) lumen.

Although the AI-based segmentation provided by PRAEVAorta is suitable for diagnostic purposes, including measurement of the AAA diameter, artifacts such as holes made it not directly applicable to our biomechanical and kinematic analyses, which require water-tight surfaces to generate patient-specific FE meshes. To facilitate the generation of such surfaces, we applied an automated post-processing pipeline to refine the segmentations obtained from PRAEVAorta. The post-processing was conducted using an in-house code developed by the Intelligent Systems for Medicine Laboratory (ISML) at The University of Western Australia (UWA). The code was implemented using MATALB programming language.

The automated post-processing consisted of the following steps:

1. Cropping the label maps to the region of interest containing the AAA.

2. Merging the three labels (aneurysm wall and ILT, calcification, and lumen) to generate a unified AAA segmentation.

3. Removing non-aortic branches and artifacts to isolate the aortic lumen.

4. Resampling both the AAA and lumen segmentations using nearest-neighbor interpolation, converting them into isotropic volumes based on the smallest voxel dimension in the original anisotropic image.

5. Initial smoothing of each segmentation to eliminate artifacts.

6. Successive refinement, including removal of surface extrusions, filling of holes, and final smoothing of each segmentation.

**Geometry**

BioPARR uses post-processed label maps and an assumed AAA wall thickness to automatically generate the AAA geometry. It creates three surfaces: the external AAA wall surface, the internal AAA wall surface, and the ILT surface.

The AAA geometry is generated in three steps. The first step is performed using 3D Slicer (https://www.slicer.org/) [3] called from within BioPARR. It involves label map manipulation to ensure that the lumen label map is fully enclosed within the AAA label map, followed by subtraction of the lumen label map from the AAA label map. A surface is then extracted using the Model Maker module in 3D Slicer. However, the initial tessellated surface generated using the Model Maker module is primarily intended for visualization and often contains poorly shaped triangles, which can hinder the analysis. To address this, in the second step, the surface is re-meshed using the surface mesh resampling software ACVD [9] to improve mesh quality. In the third step, a custom command-line interface (CLI) module in 3D Slicer is used to create the discretised (triangulated) surfaces. This module uses the re-meshed AAA surface to separate the external AAA wall surface from the ILT surface. It uses the assumed wall thickness to displace the external wall nodes along the surface normal and generate the internal AAA wall surface. The final surfaces are then exported in standard STL (stereolithography) format.

The present dataset includes one AAA geometry description for each patient: the external AAA wall surface discretised using triangles.

**Ground Truth**

We generated the synthetic ground truth image data by warping the diastolic-phase 3D-CTA image of Patient 1 using realistic wall displacements obtained from a patient-specific FE biomechanical model.

We used the AAA surface model of Patient 1, created using BioPARR (Fig. 3a), to generate a hexahedral mesh in HyperMesh (https://altair.com/hypermesh) [7], which was then imported into the Abaqus finite element (FE) software (https://www.3ds.com/products/simulia/abaqus). The mesh, shown in Fig. 3b, consisted of 20-node quadratic hexahedral elements with hybrid formulation and constant pressure, referred to as element type C3D20H in Abaqus. Boundary conditions were applied by rigidly constraining the top and bottom ends of the AAA, and a uniform pressure of 13 kPa was applied to the inner AAA wall surface.

For the wall tissue material, we adopted a hyperelastic constitutive model, with a polynomial strain energy potential given by [10]:

$$W = \alpha(I_B - 3) - \beta(I_B - 3)^2 \tag{1}$$

where $I_B$ is the first invariant of the left Cauchy–Green (Finger) deformation tensor **B**. The material parameters were taken from literature: $\alpha = 0.174$ MPa and $\beta = 1.881$ MPa [10-12]. The FE simulation was carried out using Abaqus/Standard solver. Fig. 3c shows the AAA wall displacement field obtained from the simulation, which is used as the synthetic ground truth to verify the AAA kinematics analysis method developed and used by Jamshidian et al. [1].

We used the nodal coordinates of the undeformed and deformed FE meshes with the Scattered Transform Extension [4] in 3D Slicer to create a B-spline warping transform, which was then applied to the diastolic-phase 3D-CTA image using the Transforms module in 3D Slicer to generate the synthetic systolic image.

It should be noted that, in line with FDA (Food and Drug Administration) and ASME (The American Society of Mechanical Engineers) guidelines, the plausibility of the synthetic ground truth, rather than exact biomechanical fidelity, is required for method verification [1].

# LIMITATIONS

The present dataset contains systolic and diastolic phases of 4D-CTA images, acquired at a single site (Fiona Stanley Hospital in Western Australia, Australia) for 10 AAA patients, and patient-specific AAA geometries extracted from these images. The results obtained using this dataset confirm the applicability of our automated pipeline in the patient-specific assessment of AAA wall displacements and strain [1]. However, comprehensive evaluation of the pipeline's accuracy and robustness would require application to a larger image dataset that includes images acquired at more than one site.

# ETHICS STATEMENT

The patients were recruited at Fiona Stanley Hospital in Western Australia, and informed consent was obtained prior to their participation. The study was conducted in accordance with the Declaration of Helsinki, and the protocol was approved by Human Research Ethics and Governance at South Metropolitan Health Service (HREC-SMHS) (approval code RGS3501) and by Human Research Ethics Office at The University of Western Australia (approval code RA/4/20/5913).

# CRediT AUTHOR STATEMENT

**Mostafa Jamshidian**: Conceptualisation, Methodology, Software, Data curation, Writing – original draft, Writing – review & editing.

**Adam Wittek**: Conceptualisation, Methodology, Resources, Data curation, Writing – review & editing, Project administration, Funding acquisition.

**Saeideh Sekhavat**: Methodology.

**Farah Alkhatib**: Data curation.

**Jens Carsten Ritter**: Project administration, Funding acquisition.

**Paul M. Parizel**: Project administration, Funding acquisition.

**Donatien Le Liepvre**: Resources.

**Florian Bernard**: Resources.

**Ludovic Minvielle**: Resources.

**Antoine Fondanèche**: Resources.

**Jane Polce:** Data curation, Resources.

**Christopher Wood**: Data curation, Resources.

**Karol Miller**: Conceptualisation, Methodology, Resources, Data curation, Writing – review & editing, Project administration, Funding acquisition.

# ACKNOWLEDGEMENTS

This work was supported by the Australian National Health and Medical Research Council NHMRC Ideas grant no. APP2001689.

# DECLARATION OF COMPETING INTERESTS

The authors declare that they have no known competing financial interests or personal relationships that could have appeared to influence the work reported in this paper.